\documentclass[a4paper,fleqn,usenatbib]{mnras}

\usepackage[T1]{fontenc}
\usepackage{ae,aecompl}

\usepackage{graphicx}	
\usepackage{amsmath}	
\usepackage{amssymb}	

\usepackage{color}
\usepackage{natbib}
\usepackage{graphicx}
\usepackage{lscape}
\usepackage[dvipsnames]{xcolor}

\DeclareGraphicsExtensions{.png,.pdf,.eps,.jpg,.tiff}

\newcommand{\vunit}{\mbox{\,km\,s$^{-1}$}}

\newcommand{\Msun}{\mbox{\,$M_\odot$}}
\newcommand{\Lsun}{\mbox{\,$L_\odot$}}

\newcommand{\mic}{\mbox{$\,\mu$m}} 
\newcommand{\nucl}[2]{\mbox{$^{#1}${#2}}}

\newcommand{\ltsimeq}{\raisebox{-0.6ex}{$\,\stackrel 
	{\raisebox{-.2ex}{$\textstyle <$}}{\sim}\,$}} 
\newcommand{\gtsimeq}{\raisebox{-0.6ex}{$\,\stackrel
	{\raisebox{-.2ex}{$\textstyle >$}}{\sim}\,$}} 
\newcommand{\prpsimeq}{\raisebox{-0.6ex}{$\,\stackrel
        {\raisebox{-.2ex}{$\textstyle \propto $}}{\sim}\,$}}

\newcommand{\fion}[2]{[{#1}\,{\sc {#2}}]}
\newcommand{\mon}{\mbox{V959~Mon}}
\title[Dust in the nova \mon]{A dusty rain falls on the nova V959 Monocerotis}

\author[A. Evans et al.]
{A. Evans$^1$, D. P. K. Banerjee$^2$, W. P. Varricatt$^3$, V. Joshi$^2$  \\ \\
$^{1}$Astrophysics Group, Lennard Jones Laboratory, Keele University, Keele
Staffordshire, ST5 5BG,UK\\
$^{2}$Physical Research Laboratory, Navrangpura, Ahmedabad, Gujarat
380009, India \\
$^{3}$UKIRT Observatory, Institute for Astronomy, 640 N. A'ohoku Place,
University Park, Hilo, Hawai'i 96720, U.S.A.}

\date{Accepted XXX. Received YYY; in original form ZZZ}

\pubyear{2024}

\begin{document}
\label{firstpage}
\pagerange{\pageref{firstpage}--\pageref{lastpage}}
\maketitle

\begin{abstract}
We present archival and ground-based infrared observations of the 
$\gamma$-ray-emitting
nova \mon, {covering the period 100--4205~days after the 2012 eruption.} We use these data to 
{determine that the secondary in
the nova system is a G5 main sequence star}. Data from the NEOWISE survey reveal a significant increase in
the emission at 3.4\mic\ and 4.6\mic\ at late ($\gtsimeq600$~days) times, 
which we interpret as emission by dust. Other interpretations are considered
but cannot be reconciled with the data. The presence of such late dust emission, 
and in particular its variation with time, are unprecedented in the context 
of novae. The behaviour of the dust emission suggests a qualitative interpretation
in which ejecta from the 2012 eruption encounter denser pre-eruption 
circumbinary material, giving rise to Rayleigh-Taylor instabilities that 
cause clumps of dust-bearing material to fall back towards the central binary,
the dust undergoing destruction by chemisputtering as it does so.
The observed rise in the dust temperature, the decline in the nova-dust distance
and in the dust mass, are consistent with this interpretation.
Not all novae are expected to show this behaviour, but inspection of 
resources such as NEOWISE might reveal other novae post-eruption
that do.
\end{abstract}




\section{Introduction}
Classical Novae (CNe) occur in semi-detached binary systems consisting
of a white dwarf (WD) primary and a Roche-lobe-filling secondary,
usually a main sequence dwarf \citep{CN2,basi12,woudt14}.
Material from the secondary spills onto the surface of the WD via an
accretion disc. In time the base of the accreted layer becomes 
degenerate, and a thermonuclear runaway (TNR) occurs. This is 
seen as a nova eruption, in which $10^{-6}-10^{-3}$\Msun\ of material,
enriched in metals up to Ca as a result of the TNR, is ejected at
several 100s to several 1000s of \vunit.
The WD in CN systems may be of CO type, or the more massive ONe type
\citep*{chomiuk20}

With a Galactic CN rate of $\simeq47$~year$^{-1}$, \citep{de21},
it is likely that CNe are a major source of \nucl{13}{C},
\nucl{15}{N} and \nucl{17}{O} in the Galaxy \citep[see, e.g.,][]{CN2}.
Historically, CNe have not been considered as major contributors to
the interstellar dust population \citep[see, e.g.,][]{gehrz89}
but a reanalysis by \cite{yates24} suggests that the CN contribution
may have been significantly underestimated.

\section{\mon}

\mon\ was discovered visually by S. Fujikawa on 2012 Aug 9.81. 
However, a $\gamma$-ray transient was detected by Fermi LAT 
(Fermi J0639+0548) on three consecutive days 
\citep[2012 June 22--24 (MJD 56100--56102);][]{cheung12a}, which was
subsequently associated with \mon\ by \cite{cheung12b}.
We take the date of eruption and the time origin to be MJD 56100.

\mon\ was observed with the Neil Gehrels Swift observatory 
\citep{gehrels04} shortly
after optical discovery \citep{nelson12a}. An observation
on 2012 November 18 showed that the nova had entered the 
super-soft
phase, with estimated temperature $\sim250,000\pm70,000$~K
\citep{nelson12b}. Further Swift observations followed the onset of the 
super-soft phase; the soft X-ray flux increased substantially, although
it was extremely variable \citep*{osborne12}. 
\citeauthor{osborne12} also reported a probable 7.1~hour periodicity in
the ultraviolet emission; this periodicity was confirmed using optical 
photometry by \cite*{wagner13}. The soft X-ray count rate peaked 
around 2013 January 10 (day 153), and rapidly declined
thereafter. By day 247 the super-soft phase had essentially
ended \citep{page13a}.

\cite{munari13} reported photometry of the eruption,
including the narrow-band Stromgren $b$ and $y$ bands.
They concluded that the (unobserved) visual maximum had $4<V<4.5$.
They too found the 7.1~hour periodicity, which they ascribed
to orbital modulation arising from a combination of ellipsoidal
variations and irradiation of the secondary by the still-hot WD.
Using their optical photometry and data from 
the Two Micron All Sky Survey \citep[2MASS;][]{2mass}, 
and assuming a 
distance 1.5~kpc and reddening $E(B-V)=0.38$, \citeauthor{munari13}
determined that the secondary is an early K~main sequence star.
Such a star would fill its Roche lobe for a 7.1~hour orbital period
and an ONe WD.

Extensive optical spectroscopy of \mon\ was reported by \cite{shore13}.
They found that the ejecta had axi\-symmetric conical/bipolar geometry,
the inclination of the axis to the line of sight, $i$, lying in the range 
$60^\circ \le i \le 80^\circ$. The half-width at zero intensity of
the emission lines was $\sim2000$\vunit. They deduced that the 
reddening to the nova is $E(B-V) = 0.85\pm0.05$. They too concluded that
\mon\ was an ONe CN, with similarites to other ONe novae like
V1974~Cyg and V382~Vel. They determined an ejecta mass $\le6\times10^{-5}$\Msun.
They also suggested that $\gamma$-ray emission
could be a feature of all ONe novae, possibly even of all CNe.

\cite*{ribeiro13} used the \fion{O}{iii} 4959, 5007\AA\ lines to 
investigate the morphology of the ejecta. They determined that the 
ejecta had bipolar structure with inclination angle of $82^\circ\pm6^\circ$ 
and had maximum expansion velocity $\simeq2400$\vunit\ on day 130. 
{\mon\ was imaged with WFC3 on the
Hubble Space telescope 882 and 1256~days into the
eruption \citep{sokoloski16}. These observations
confirmed the bipolar structure, which had major axis 
1\farcs05 on day 882, 
consistent with the morphology deduced by
\citeauthor{ribeiro13}.}

Near infrared (NIR) spectra obtained by \cite*{banerjee12} on 
2012 November 1--2 showed the presence of the coronal lines
\fion{S}{ix} 1.252\mic, \fion{Si}{vi} 1.964\mic, \fion{Al}{ix} 2.040\mic,
with other coronal lines suspected.

Radio observations of \mon\ have been described by \cite{healy17}
and \cite{linford15}. \citeauthor{healy17} found that, while the 
source was initially elongated E--W, it later became elongated N--S.
\citeauthor{linford15} also monitored the evolution of the ejecta
morphology. They combined the expansion of the radio remnant
with optical spectroscopy to estimate a distance in the range
$0.9\pm0.2$~kpc to $2.2\pm0.4$~kpc, with a most probable value of 
$1.4\pm0.4$~kpc, consistent with the \cite{munari13} value.
The Gaia3 geometric distance is $D=2.7$~kpc, the photo-geometric distance is 
3.0~kpc \citep{bailer21}. Using the [3D] Galactic extinction map based on 
Gaia parallaxes \citep{green19}, the reddening
$E(g-r)$ is $0.46^{+0.04}_{-0.03}$ for the lower distance, 
and $0.71^{+0.06}_{-0.03}$ for the higher. 

In this paper we present ground-based and archival infrared (IR) 
data on the CN \mon. We take 3~kpc for the distance. 
Assuming that $E(B-V) = 0.981 E(g-r)$ \citep{schlafly11}, 
the corresponding reddening is $E(B-V)=0.7$, which we also assume.
The data have been dereddened using the reddening law given by
\cite*{cardelli89},

\section{The data}

\subsection{Near infrared photometry}

\begin{figure}
\centering
 \includegraphics[width=8cm,keepaspectratio]{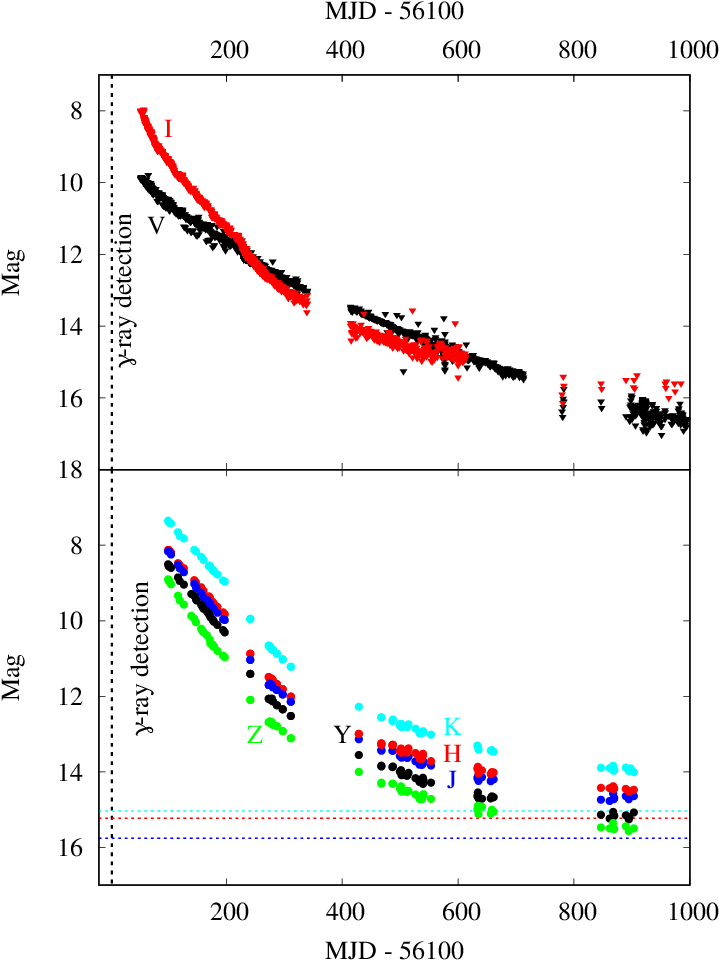}
 \caption{Top: $V$ and $I$ light curves from AAVSO data.
 Bottom: UKIRT photometry {covering the same
 time-frame as the AAVSO data}; the horizontal lines denote
 the pre-eruption values from the 2MASS survey.
 In both cases the time of the first $\gamma$-ray detection (see text)
 is indicated by the dashed vertical lines.
 \label{LC}}
\end{figure}

\begin{table*}
\caption{Portion of UKIRT photometry. The entire dataset is available online.
\label{ukirt}}
 \begin{tabular}{ccccccc}   \hline
UT Date  &  UT     &  MJD & {Days past} &Filter &   Magnitude   & Magnitude \\
yyyymmdd  &  hh:mm:ss  &    &  {eruption} &      &    & error (1$\sigma$) \\ \hline
&&&&&&\\
20120929  &  14:31:29  &  56199.60520  &   99.60520  &  $K$  &  7.350 &   0.005 \\
20120929  &  14:34:03  &  56199.60698  &   99.60698  &  $H$  &  8.127 &   0.003 \\
20120929  &  14:36:32  &  56199.60870  &   99.60870  &  $J$  &  8.166 &   0.008 \\
20120929  &  14:39:02  &  56199.61044  &   99.61044  &  $Y$  &  8.508 &   0.010 \\
20120929  &  14:41:45  &  56199.61233  &   99.61233  &  $Z$  &  8.904 &  0.014 \\
&&&&&&\\
20120930  &  13:17:45  &  56200.55399  &  100.55399  &  $K$  &  7.384 &  0.012 \\
20120930  &  13:20:38  &  56200.55600  &  100.55600  &  $H$  &  8.125 &   0.017 \\
20120930  &  13:23:14  &  56200.55780  &  100.55780  &  $J$  &  8.165 &   0.007 \\
20120930  &  13:25:51  &  56200.55962  &  100.55962  &  $Y$  &  8.562 &   0.029 \\
20120930  &  13:28:33  &  56200.56149  &  100.56149  &  $Z$  &  8.948 &   0.013 \\
&&&&&&\\
20121001  &  13:21:48  &  56201.55681  &  101.55681  &  $K$  &  7.404 &   0.019 \\
20121001  &  13:24:18  &  56201.55854  &  101.55854  &  $H$  &  8.151 &   0.013 \\
20121001  &  13:26:55  &  56201.56036  &  101.56036  &  $J$  &  8.174 &   0.012 \\
20121001  &  13:29:25  &  56201.56209  &  101.56209  &  $Y$  &  8.536 &   0.011 \\
20121001  &  13:32:07  &  56201.56397  &  101.56397  &  $Z$  &  8.942 &   0.011 \\
&&&&&&\\
20121004  &  14:35:28  &  56204.60796  &  104.60796  &  $K$  &  7.427 &   0.029 \\
20121004  &  14:36:28  &  56204.60866  &  104.60866  &  $H$  &  8.190 &   0.009 \\
20121004  &  14:37:33  &  56204.60941  &  104.60941  &  $J$  &  8.248 &   0.018 \\
20121004  &  14:38:33  &  56204.61010  &  104.61010  &  $Y$  &  8.600 &   0.013 \\
20121004  &  14:39:33  &  56204.61080  &  104.61080  &  $Z$  &  9.035 &   0.015 \\
&&&&&&\\
20121016  &  15:22:05  &  56216.64034  &  116.64034  &  $K$  &  7.658 &   0.022 \\
20121016  &  15:23:04  &  56216.64102  &  116.64102  &  $H$  &  8.475 &   0.018 \\
20121016  &  15:24:03  &  56216.64170  &  116.64170  &  $J$  &  8.520 &   0.011 \\
20121016  &  15:25:04  &  56216.64241  &  116.64241  &  $Y$  &  8.855 &   0.012  \\
20121016  &  15:26:03  &  56216.64309  &  116.64309  &  $Z$  &  9.336 &   0.006 \\
&&&&&&\\            
20121019  &  13:19:43  &  56219.55536  &  119.55536  &  $K$  &  7.758  &  0.018 \\
20121019  &  13:20:44  &  56219.55606  &  119.55606  &  $H$  &  8.523  &  0.013 \\
20121019  &  13:21:43  &  56219.55675  &  119.55675  &  $J$  &  8.614  &  0.007 \\
20121019  &  13:22:42  &  56219.55743  &  119.55743  &  $Y$  &  8.949  &  0.014 \\
20121019  &  13:23:43  &  56219.55814  &  119.55814  &  $Z$  &  9.462  &  0.013 \\
&&&&&&\\
\multicolumn{6}{l}{Continued online.} \\
\hline
\end{tabular}
\end{table*}

Observations of \mon\ were obtained with the 3.8-m United Kingdom Infrared 
Telescope (UKIRT) and the Wide Field Camera \citep[WFCAM;][]{casali07}, 
using the near-IR MKO filters $Z, Y, J, H$ and $K$ (effective wavelengths: 
0.88, 1.03, 1.25, 1.63 and 2.20\mic\ respectively). WFCAM has a pixel 
scale of $0\farcs4$/pix and employs four $2048\times2048$ HgCdTe Hawai'iII
arrays. Each array has a field of view of $13\farcm65\times13\farcm65$.
Observations were performed  by locating the object in one of the four
arrays and by dithering to five points separated 
by a few arcseconds. 
{For all but the very first three epochs
(for which the exposure per filter was 40~seconds), the total
on-chip exposure was 10 seconds per filter.} 
The data were reduced by the Cambridge Astronomical Survey Unit 
(CASU); the archiving and distribution of the data are carried out by the 
Wide Field Astronomy Unit (WFAU).

The monitoring observations of the nova were carried out as a backup
program, 
so the observations were sometimes obtained in the presence of clouds.  
{The magnitudes of the nova from observations
during the period up to the end of 2013 were}
calculated using the average of the zero points for a set of
eight 
isolated point sources present in all dithered frames around the nova, 
and using their magnitudes measured on nights when the sky was photometric.
The errors listed in Table~\ref{ukirt} are the 1-$\sigma$ of the zero point 
estimates for these objects. 
Most of these observations were obtained with the telescope kept out of focus 
to avoid saturation, so we have used a $12''$-diameter aperture for photometry. 
Photometry was perfomed using the Starlink task ``Autophotom''.
{The observations from 2014, when the source
became faint, were carried out with the telescope in focus.
The magnitudes reported are extracted from the catalogues
produced by the data processing by CASU, with the zero points
estimated from isolated point sources in the field.}

NIR photometry from 2012 September 29 to October 26, and from 2012 
November 8 to 2013 May 5, were reported by 
\cite{varricatt12a,varricatt12b} and are included in this paper for
completeness. The magnitudes for the first twenty days of observation, 
and the  UT and MJD of mid-observation, are given in Table~\ref{ukirt}.
The complete dataset,
{which covers the period from 100 to 4206~days
after the 2012 eruption,}
is available online. 

The UKIRT photometry is shown, along with $V$ and $I$ data from 
the AAVSO\footnote{https://www.aavso.org/} database, in Fig.~\ref{LC}.
{This figure shows UKIRT data only to day
1000 after the eruption, 
when the AAVSO data are available.}

\subsection{WISE and NEOWISE}
The Wide-field Infrared Survey Explorer \citep[WISE;][]{wise} conducted 
an all-sky
survey in wavebands centred on 3.4\mic\ (W1), 4.6\mic\ (W2), 12\mic\ (W3), 
and 22\mic\ (W4). \mon\ was detected in WISE bands W1 and W2 
in 2010 March/October with mean
fluxes $0.412\pm0.012$~mJy and $0.186\pm0.013$~mJy respectively
\citep{evans14}.

The prime science driver of the Near-Earth Object + WISE 
\citep[NEOWISE;][]{neowise,neowise2} surveys was the identification
of moving (solar system) objects detected in the WISE survey.
The NEOWISE Reactivation Mission \citep{neowise2} provided data in
WISE bands W1 and W2 only.

\begin{figure}
\centering
\includegraphics[width=5.5cm,angle=180]{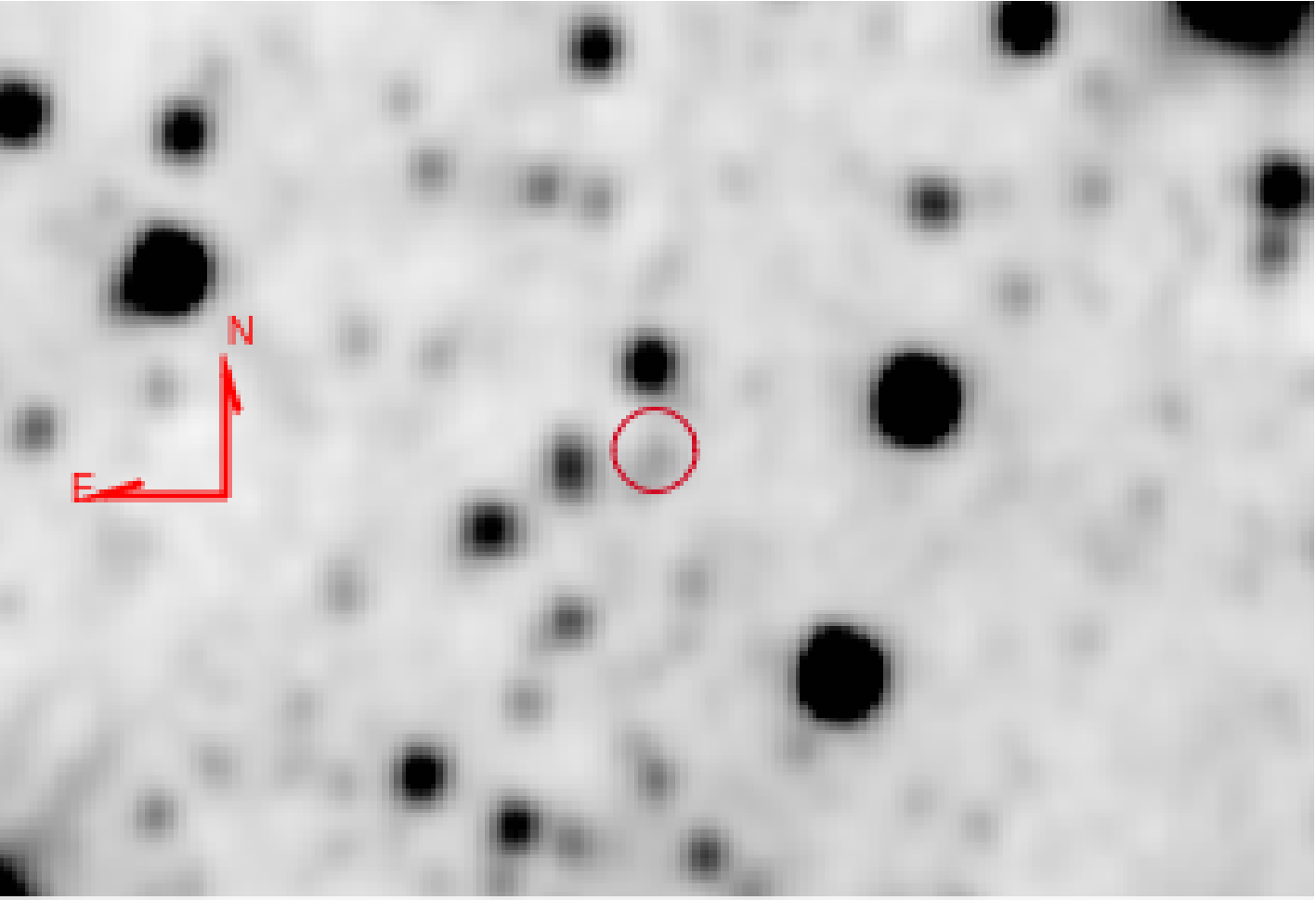}
\includegraphics[width=5.5cm]{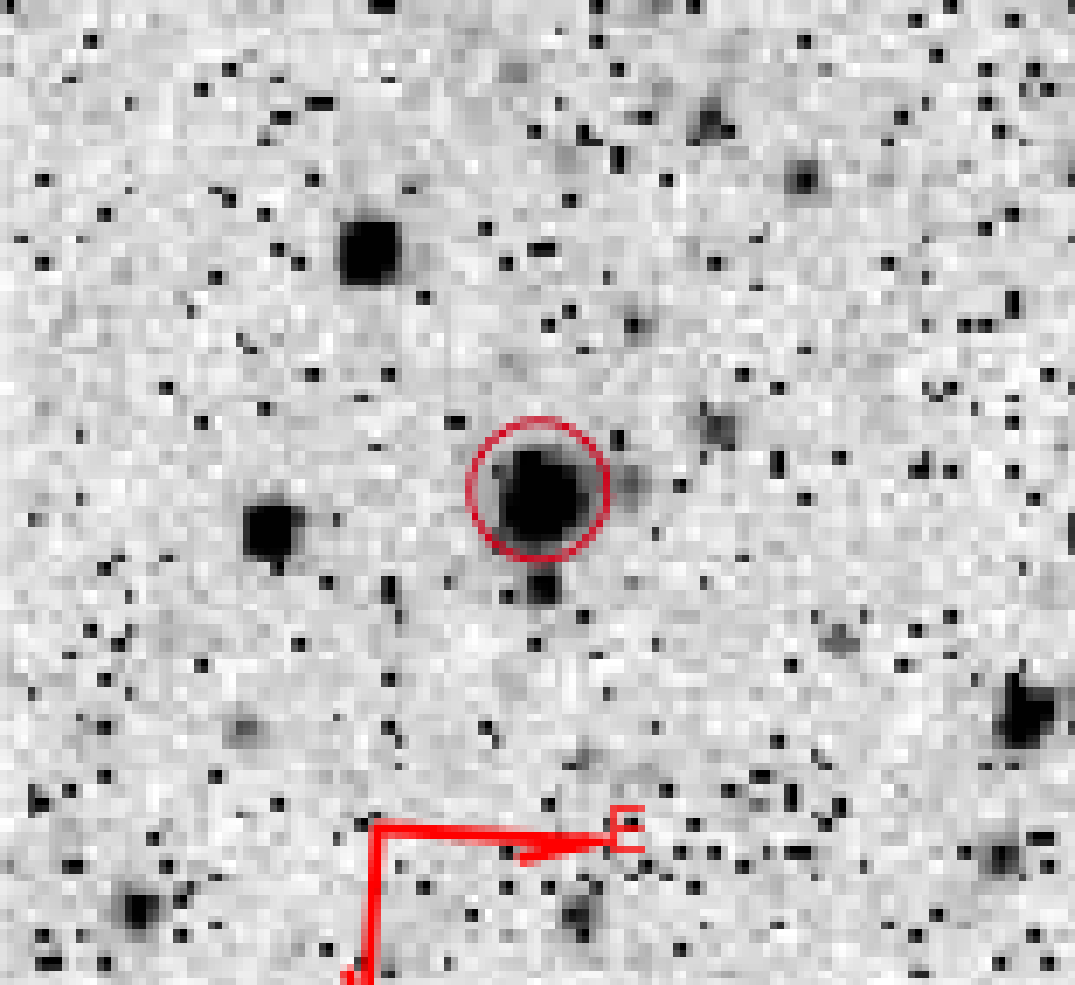}
\includegraphics[width=5.5cm]{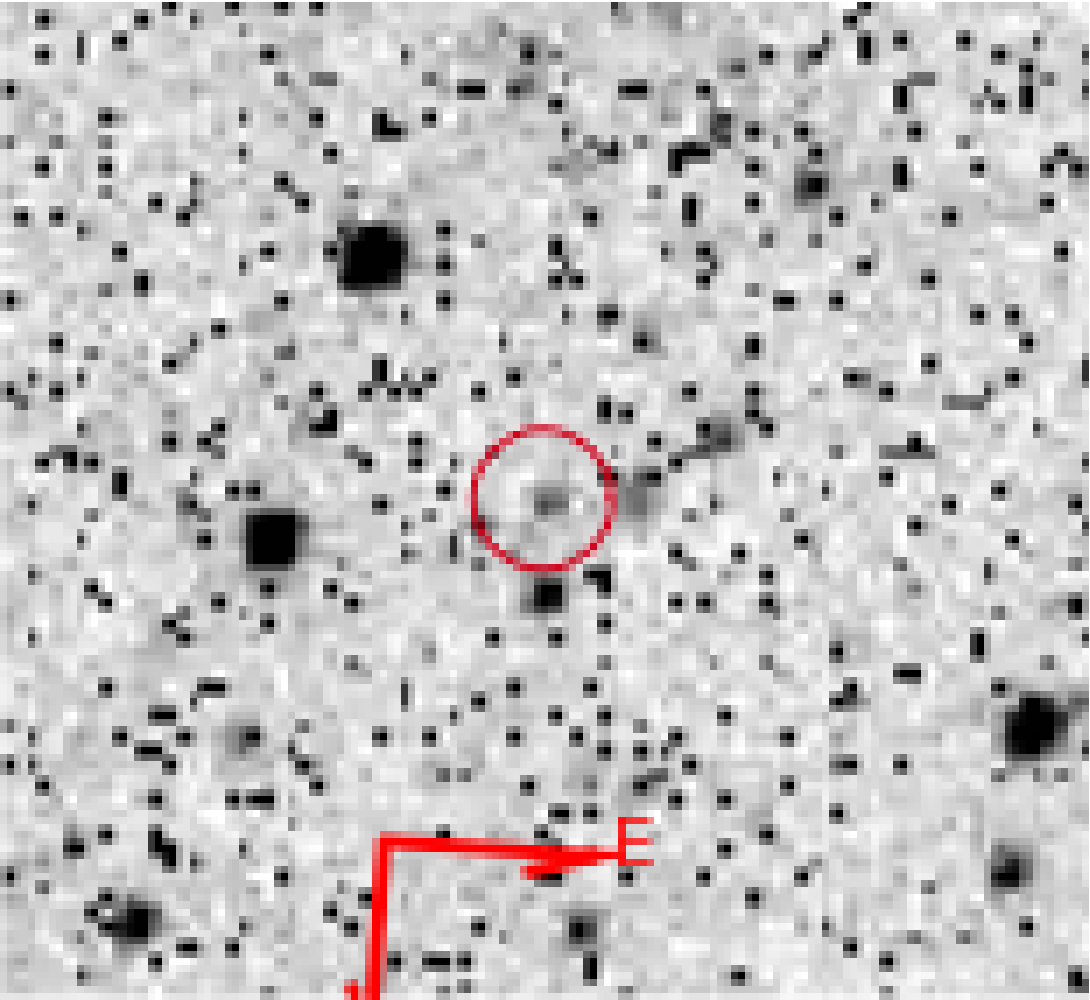}
\includegraphics[width=5.5cm,angle=180]{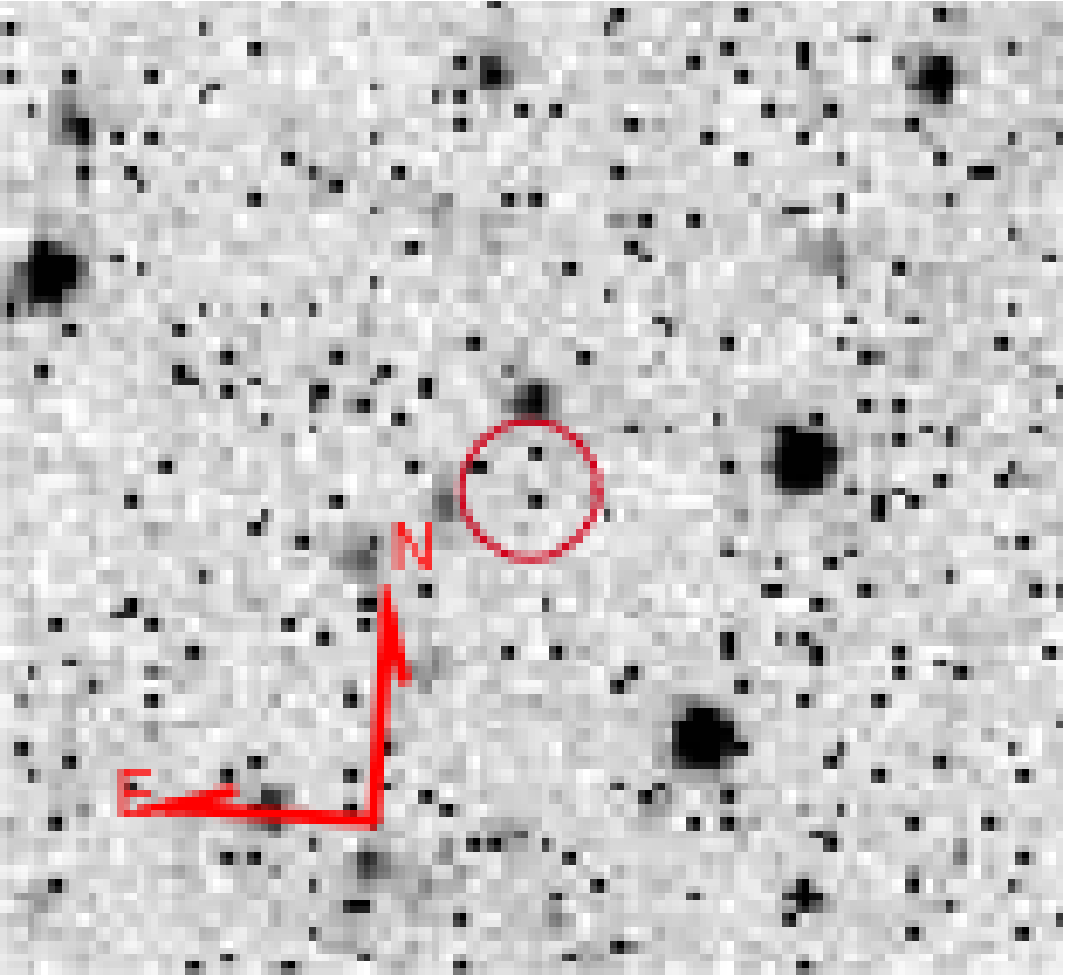}
 \caption{WISE and NEOWISE W2 images of \mon\ obtained on MJD 55278 (WISE; top, 822~days before eruption), on MJD 56753 (NEOWISE; 
 middle top, 643~days after eruption, around the
 time of the peak in the W2 flux), on MJD 58186 (middle bottom, 2086~days after eruption, during W2 declie) and 59855 (bottom, 3755 days after eruption, after the W2 flux had faded).
 \mon\ is marked with the red circle. WISE image is $5'$ square,
 NEOWISE images are $4'$ square.\label{W2}}
\end{figure}

We have trawled the NEOWISE data for \mon\ and it is clearly detected 
in bands W1 and W2 at fluxes that far exceed the pre-outburst values 
given in \cite{evans14} (see Fig.~\ref{W2}). 
{The positions of the NEOWISE sources agree with that
of \mon\ within the spatial resolution of the survey.
Moreover, the variation of the NEOWISE source makes the 
identification with the nova certain.}

\begin{figure}
 \includegraphics[width=8cm,keepaspectratio]{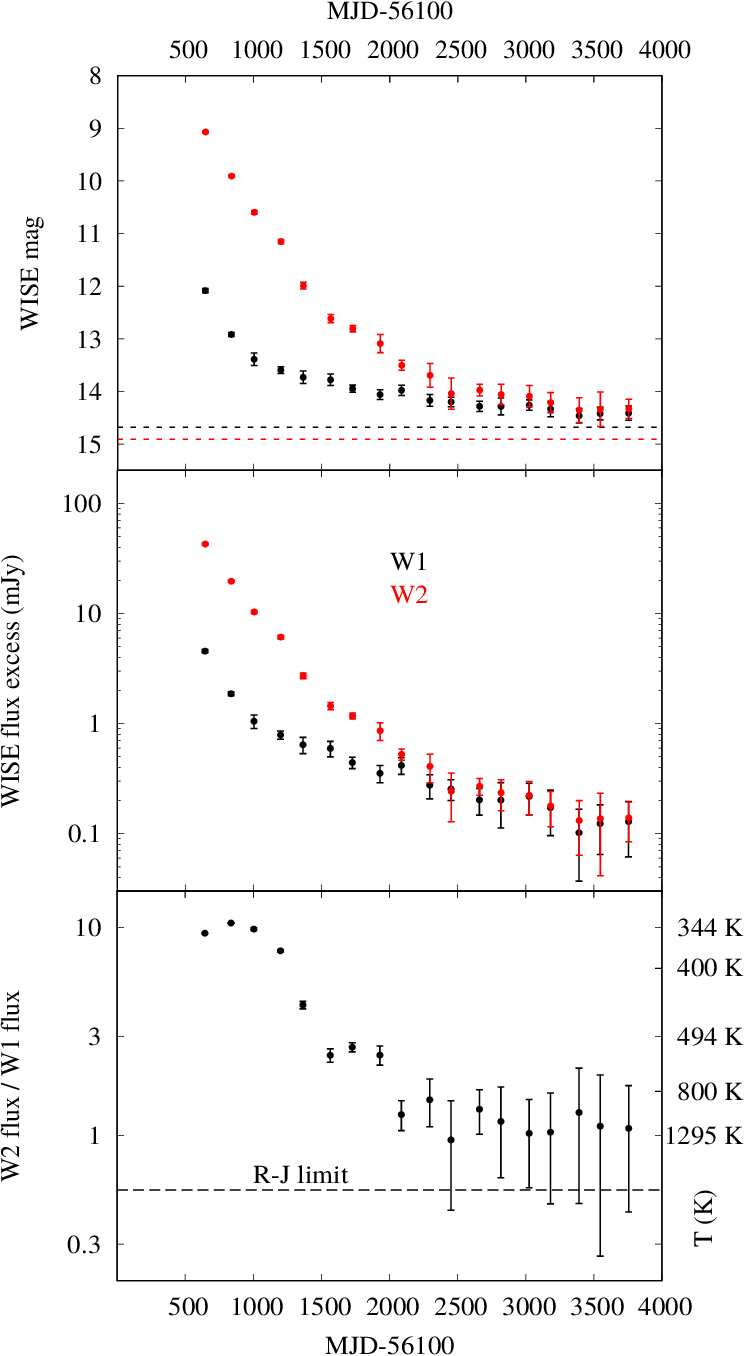}
 \caption{Top: NEOWISE W1 and W2 light curves.
 Horizontal lines denote pre-outburst values from WISE survey.
 Middle: Variation of the flux from the 2012 ejecta, i.e.
 {the quiescent WISE fluxes have been
 subtracted from the NEOWISE data; the excess fluxes have been} dereddened by $E(B-V)=0.7$.
 Bottom: Variation of the flux excess ratio W2/W1; 
 flux excesses are dereddened. The error bars are derived from the errors in the W1 and W2 fluxes. The temperature scale on the 
 right-hand axis corresponds to that of the black body having the W2/W1 
 flux ratio on the left-hand axis.
 {The broken horizontal line is the
 Rayleigh-Jeans limit.}
 \label{wise_lc}}
\end{figure}

The nature of the NEOWISE 
survey means that the data consist of closely-spaced ``blocks'' of 
several observations (typically covering $\sim0.3-2$~days), obtained 
within a very short time of each other. 
These data have been averaged to provide a mean value for the block
and are given in Table~\ref{neowise}.
The W1 and W2 light curves are shown in Fig.~\ref{wise_lc}.

\begin{table}
\caption{NEOWISE photometry, in WISE magnitudes.\label{neowise}}
\begin{tabular}{ccc}
{Days past eruption} & W1 &   W2 \\
(MJD--56100$^*$) & (3.4\mic) &  (4.6\mic) \\ \hline
 644.2827 & $12.0840\pm0.0380$ &  $9.0683\pm0.0219$  \\
 835.8767 & $12.9166\pm0.0382$ &  $9.9066\pm0.0228$  \\
1003.2360 & $13.3875\pm0.1177$ & $10.5955\pm0.0360$  \\
1199.0230 & $13.5931\pm0.0638$ & $11.1521\pm0.0407$  \\
1363.8035 & $13.7298\pm0.1181$ & $11.9879\pm0.0635$  \\
1564.8266 & $13.7779\pm0.1094$ & $12.6134\pm0.0771$  \\
1726.6757 & $13.9470\pm0.0689$ & $12.8065\pm0.0606$  \\
1929.0897 & $14.0597\pm0.0933$ & $13.0896\pm0.1715$  \\
2086.4785 & $13.9777\pm0.0970$ & $13.5023\pm0.0948$  \\
2295.1683 & $14.1693\pm0.1107$ & $13.6934\pm0.2260$  \\
2450.6840 & $14.2005\pm0.0901$ & $14.0413\pm0.2972$  \\
2660.2145 & $14.2830\pm0.0980$ & $13.9743\pm0.1117$  \\ 
2817.7376 & $14.2841\pm0.1617$ & $14.0583\pm0.1943$  \\
3025.0720 & $14.2582\pm0.0974$ & $14.0904\pm0.2027$  \\
3181.8622 & $14.3334\pm0.1446$ & $14.2110\pm0.1900$  \\ 
3391.6794 & $14.4616\pm0.1377$ & $14.3551\pm0.2341$  \\
3546.4362 & $14.4208\pm0.1201$ & $14.3377\pm0.3272$  \\
3755.7894 & $14.4113\pm0.1361$ & $14.3302\pm0.1828$  \\ \hline\hline
\multicolumn{3}{l}{$^*$Mean MJD after averaging over a block, as described
in text.}\\
\multicolumn{3}{l}{Spread in mean JD is in the range $0.3-2$~days.}\\
\end{tabular}
\end{table}

\subsection{Herschel}
\begin{figure}
\centering
\includegraphics[width=4.5cm,keepaspectratio,angle=-90]{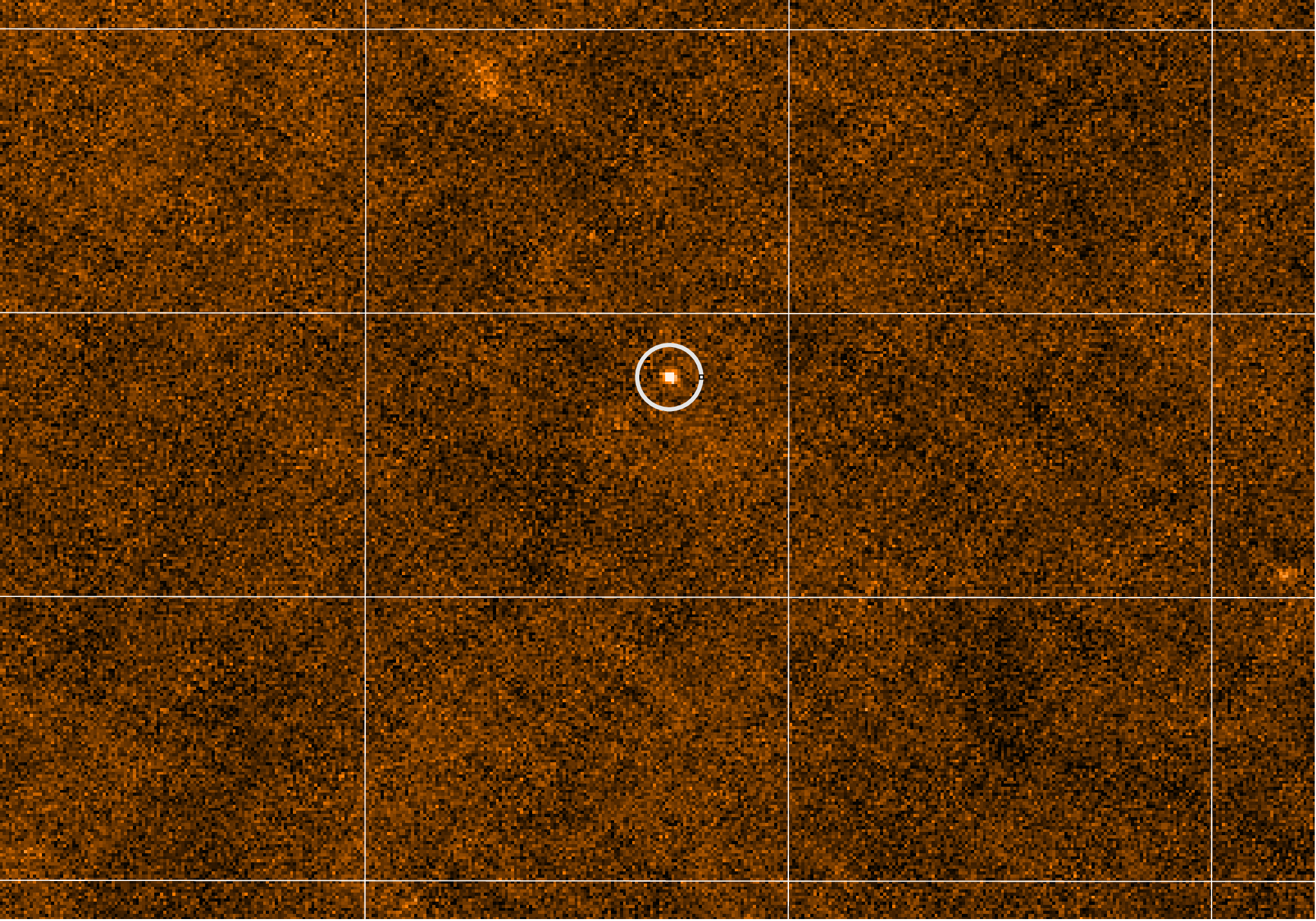}
\includegraphics[width=4.5cm,keepaspectratio,angle=-90]{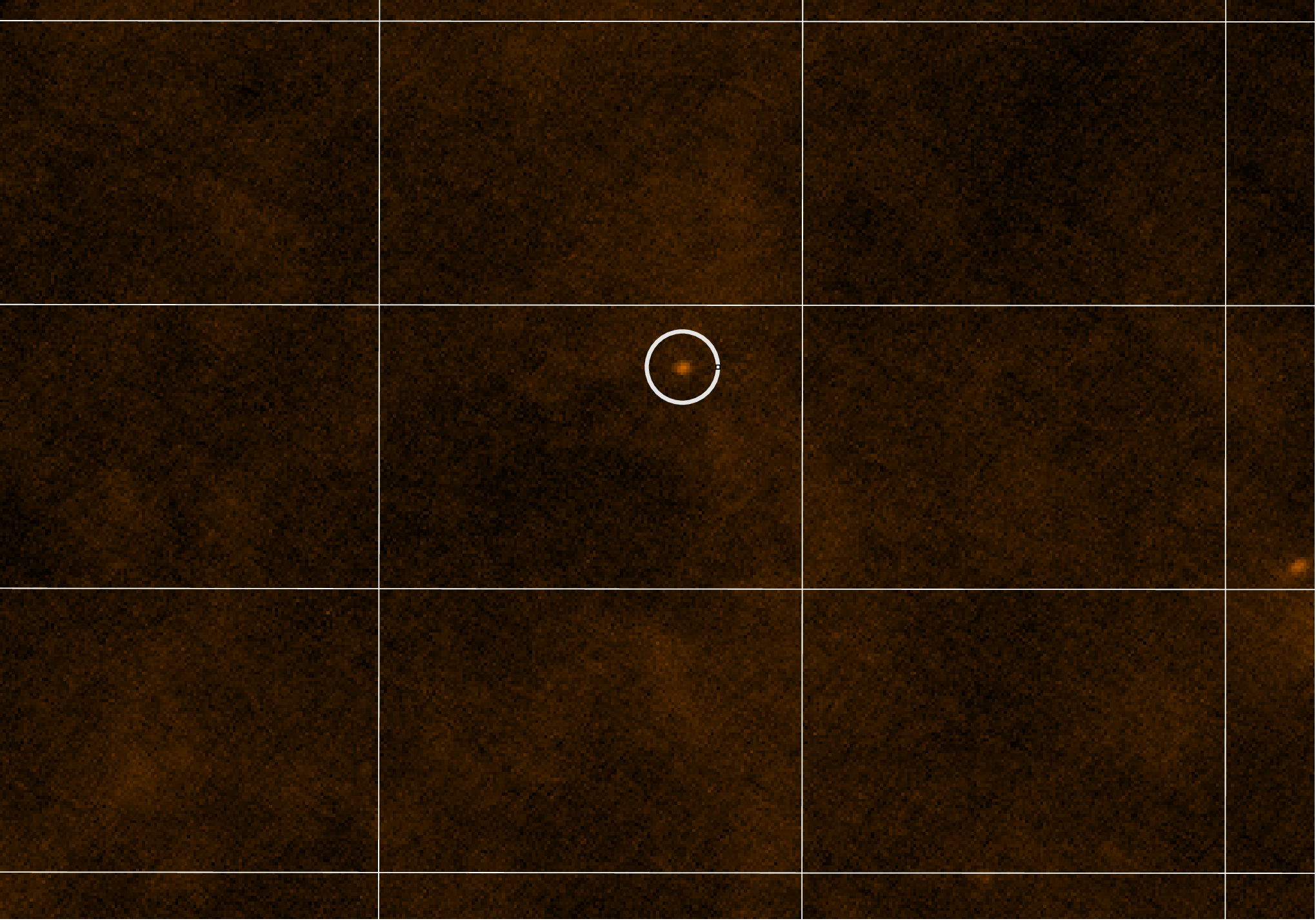}
 \includegraphics[width=4.5cm,keepaspectratio,angle=-90]{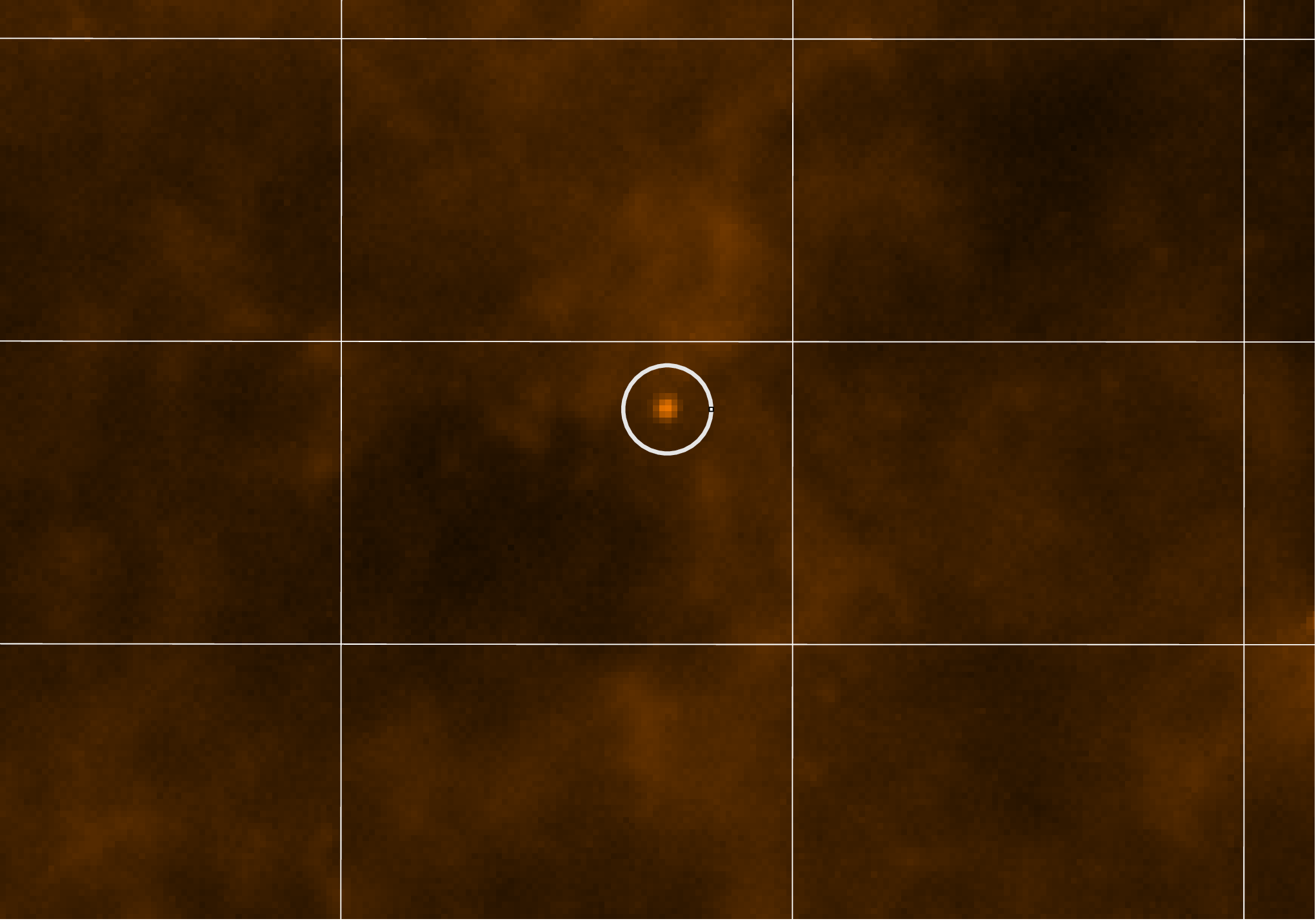}
 \includegraphics[width=4.5cm,keepaspectratio,angle=-90]{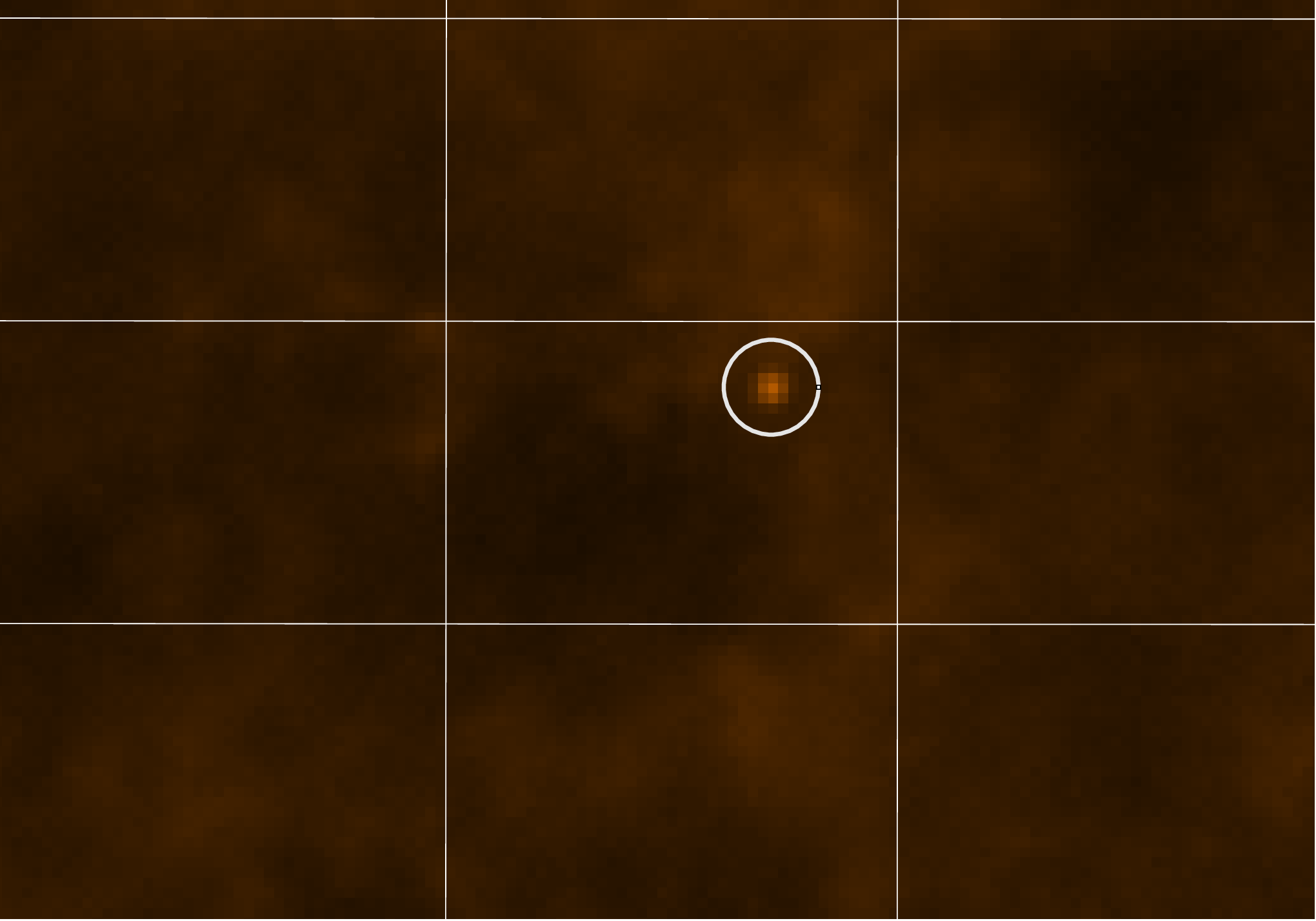}
 \includegraphics[width=4.5cm,keepaspectratio,angle=-90]{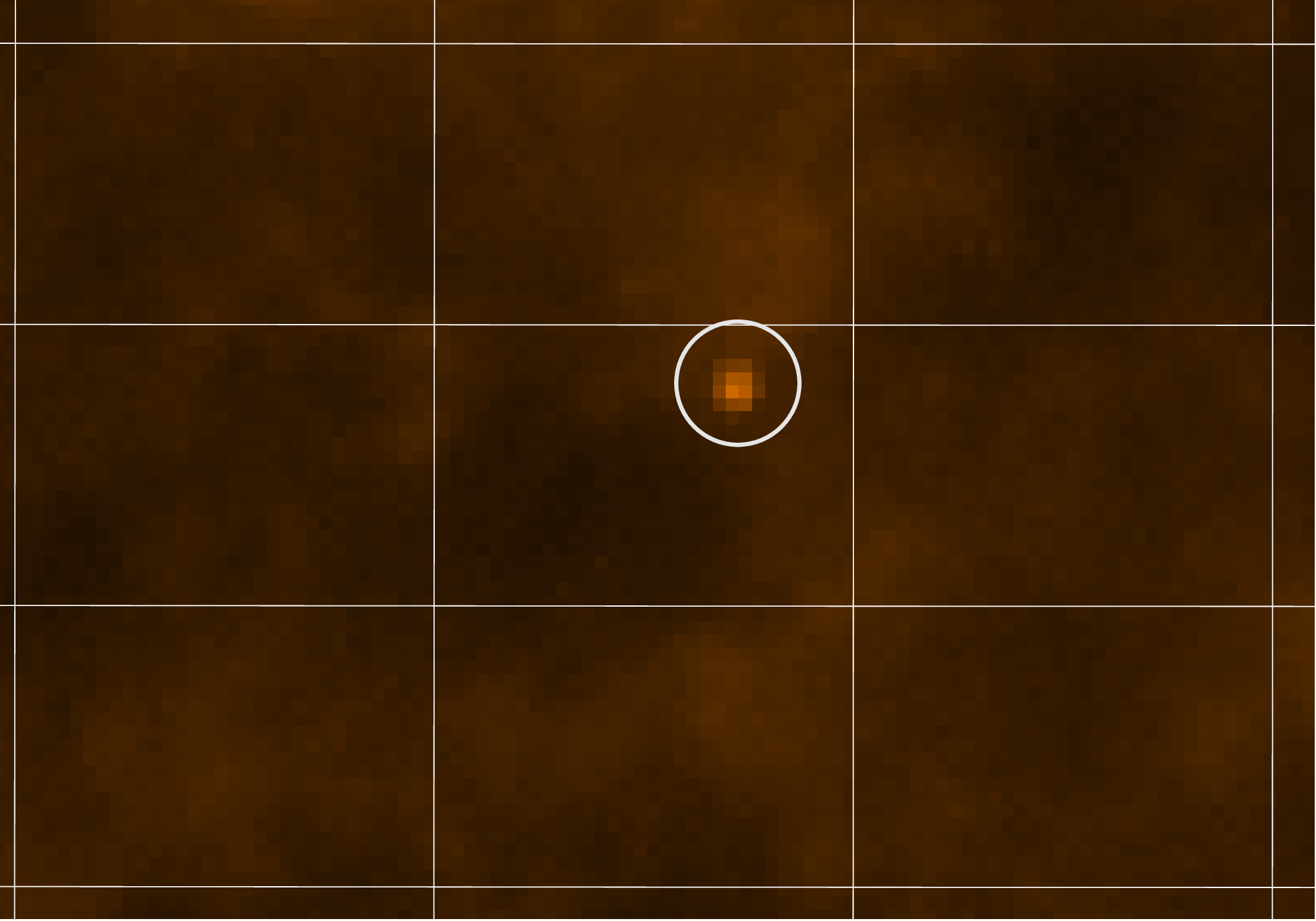}
 \caption{From top to botton, Herschel 70\mic, 160\mic, 250\mic,
 350\mic, 500\mic\ images of \mon; the nova is 
 at the centre of the circle. In each frame north is up, east is
 left. The rectangles are $7\farcm5\times5'$ in RA$\times$Dec.
 \label{herschel-f}}
\end{figure}

We have also trawled the data in the Photodetector Array Camera and
Spectrometer \citep[PACS;][]{poglitsch10} and Spectral and Photometric Imaging
REceiver \citep[SPIRE][]{spire} instruments  on the 
Herschel Space Observatory \citep{pilbratt03,pilbratt10}.
The region around \mon\ was observed on 12 October 2012
(MJD 56212), on day~112 of the eruption and the data clearly show that
\mon\ was detected. 
{As with the NEOWISE data, the position of the 
Herschel source agrees, within the spatial resolution,
with that of \mon.}
Images from the Herschel archive are shown in Fig.~\ref{herschel-f}
and the fluxes are given in Table~\ref{herschel-t}.

\begin{table}
\caption{Herschel PACS and SPIRE photometry.\label{herschel-t}}
 \begin{tabular}{ccl}
$\lambda$ (\mic) & $f_\nu$ (mJy) & \multicolumn{1}{c}{Instrument} \\ \hline
 70 &   $669.565\pm16.905$  &  PACS \\
160 &   $721.850\pm66.152$    &  PACS\\
250 &  $795.6\pm74.5$      &  SPIRE PSC\\
350 &  $864.4\pm87.8$      & SPIRE PSC\\
500 &  $876.9\pm78.2$      & SPIRE PSC\\ \hline
 \end{tabular}
\end{table}

\section{Interpreting the data}

\subsection{The progenitor}

\begin{figure}
\centering
 \includegraphics[width=8cm,keepaspectratio]{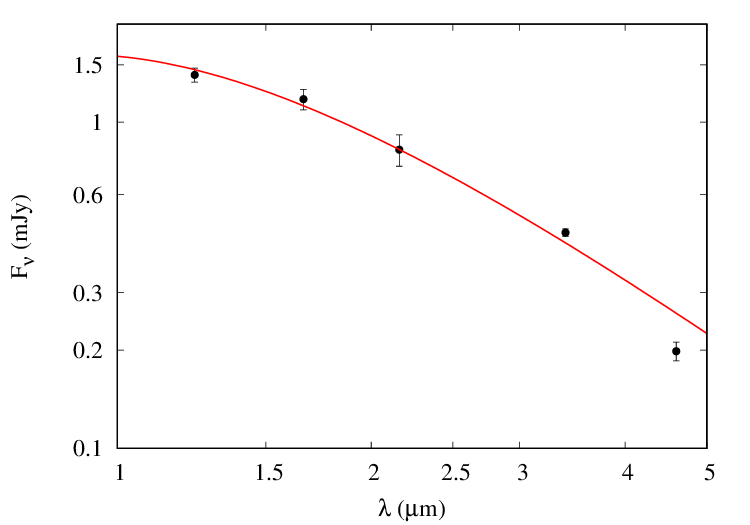}
 \caption{Pre-outburst NIR SED for \mon. Data dereddened as discussed
 in text. Curve is a weighted fit of a 5660~K black body to the data.
 See text for details.
 \label{preob}}
\end{figure}

The WISE data were obtained before the 2012 eruption of \mon, and have 
been combined with pre-eruption data from the 2MASS survey. The dereddened
data are shown in Fig.~\ref{preob}. We use these data to determine the nature
of the progenitor.

The data are fitted by a black body with temperature 
$T_{\rm prog} = 5660\pm950$~K. This black body has 
$[\lambda{f}_{\lambda}]_{\rm max}=6.40\times10^{-15}$~W~m$^{-2}$,
or $L_{\rm bol}=2.44$\Lsun\ at 3~kpc. These properties are consistent with a
G5 main sequence star. \cite{munari13} concluded that the secondary 
is a K3 main sequence star, based on reddening $E(B-V)=0.30$ and
distance 1.5~kpc. The greater reddening assumed here leads to a higher 
effective temperature than that found by \citeauthor{munari13}. 

\subsection{The Herschel data\label{herschel}}

\begin{figure}
 \includegraphics[width=8cm,keepaspectratio]{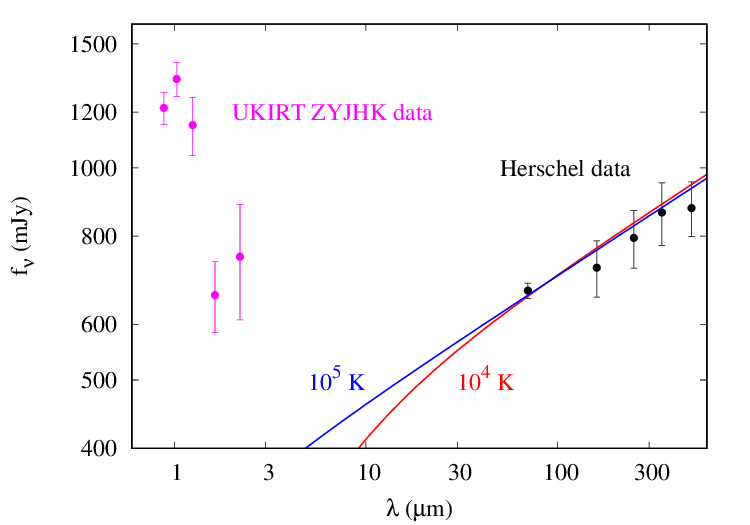}
 \caption{Fit of optically thin free-free emission
 to Herschel data at the temperatures indicated. Near-simultaneous
 UKIRT $ZY\!JH\!K$ photometry is also included. See text for details. 
 \label{herschel-f2}}
\end{figure}

The Herschel photometry is plotted in Fig.~\ref{herschel-f2}. 
There are UKIRT WFCAM data a few days on either side of the Herschel
observation. These have been interpolated to the time of the Herschel
observation, and are also plotted in Fig.~\ref{herschel-f2}. 
The dependence of $f_\nu$ on $\lambda$ at the
longest (Herschel) wavelengths suggests that we have
optically thin free-free emission, for which the luminosity is 
\citep[see, e.g.,][]{AQ}
\begin{equation}
 L_{\rm ff} = \mbox{Const.~} \times T_{\rm gas}^{-1/2} Z^2 n_e n_i V g_{\rm ff} \exp[-hc/\lambda{k}T_{\rm gas}] \:\:. \label{fff}
 \end{equation}
Here ``Const.'' is a combination of fundamental constants, $T_{\rm gas}$ 
is the plasma temperature, $Z$ is the ionic charge, $n_e, n_i$ are the 
electron and ion densities respectively, $V$ is the emitting volume and
$g_{\rm ff}$ is the free-free gaunt factor \citep{karzas61}.
In addition to the dependence $f_\nu\propto\lambda^{0}$ for optically thin
free-free emission, the dependence of $f_\nu$ on $\lambda$ arises primarily 
from the gaunt factor $g_{\rm ff}$, which over the wavelength range and likely
temperature range of interest here has the approximate form 
$g_{\rm ff} \prpsimeq \lambda^{0.17}$ \citep[from fitting data in][]{karzas61}. 
We therefore fit the data using 
$f_\nu = A\:\lambda^{0.17}\exp[-hc/\lambda{k}T_{\rm gas}]$, 
with $T_{\rm gas}=10^4$~K and $T_{\rm gas}=10^5$~K, to give (for 
$\lambda$ in \mic)
\begin{eqnarray*}
 f_\nu \mbox{~(in mJy)} &=& 3.16[\pm0.05]\times10^5 \lambda^{0.17} \exp[-hc/\lambda{k}T_{\rm gas}] \\
   && \mbox{~~~~~~~~~~~~~~~~~~~~~~~~(for $T_{\rm gas}=10^4$~K)}  \\
  f_\nu \mbox{~(in mJy)}&=& 9.86[\pm0.13]\times10^5  \lambda^{0.17} \exp[-hc/\lambda{k}T_{\rm gas}]  \\
  & & \hfill \mbox{~~~~~~~~~~~~~~~~~~~~~~~~(for $T_{\rm gas}=10^5$~K)} \:\:.
\end{eqnarray*}

These fits are shown in Fig.~\ref{herschel-f2}. At short wavelengths
the continuum
for the $T_{\rm gas}=10^4$~K case starts to turn down as a result of the 
$\exp[-hc/\lambda{k}T_{\rm gas}]$ term. It is evident
that the free-free emission makes negligible contribution
to the emission at the 
shorter (UKIRT $Y\!Z\!J\!H\!K$) wavelengths. It is also evident that the plasma 
temperature is not well constrained. 

From the fit we can determine the emitting 
mass, assuming pure hydrogen ($Z=1$). We find
\begin{equation}
 \frac{M_{\rm H}}{\Msun} \simeq \frac{\mbox{const}}{n_e\mbox{~(in cm$^{-3}$)} } \left(\frac{D}{\mbox{3~kpc}}\right)^2\:\:, \label{m-ff}
\end{equation}
where const $=3.24\times10^5$ for the  $T_{\rm gas}=10^4$~K 
case, and const $=3.19\times10^6$ for the  $T_{\rm gas}=10^5$~K case.
For higher (lower) values of $T_{\rm gas}$, the turn down shifts to 
shorter (longer) wavelengths. The \fion{Si}{vi} 1.96\mic\ line (critical 
density $\sim4\times10^6$~cm$^{-3}$ at $T_{\rm gas}=10^4$~K,
$\sim1.3\times10^7$~cm$^{-3}$ at $T_{\rm gas}=10^5$~K) was observed
within $\sim30$~days of the Herschel observation \citep{banerjee12}.
If the \fion{Si}{vi} line and free-free emission both arose in the 
same region of the ejecta, the electron density in the \fion{Si}{vi}-emitting
region was less than these values on 2012 October 12. The emitting mass is 
then $\gtsimeq0.081$\Msun\ ($\gtsimeq0.079$\Msun) for $T_{\rm gas}=10^4$~K
($T_{\rm gas}=10^5$~K). So $M_{\rm H}\gtsimeq0.08$\Msun\ irrespective of 
the value of $T_{\rm gas}$.

An alternative estimate for the electron density is given 
by \cite{shore13}, who used the \fion{O}{iii} 4636\AA\ line to conclude that 
$n_e\simeq3\times10^7$~cm$^{-3}$ on 2012 November 21 (day~152, 
40 days later than the Herschel observation), 
for electron temperature $10^4$~K. Assuming that the density declines
as $t^{-2}$, the electron density on day 112 would have been
$5.5\times10^7$~cm$^{-3}$, leading to 
$M_{\rm H}/\Msun\simeq6\times10^{-3}$.

These estimates of $M_{\rm H}$ are
significantly larger than the ejecta 
mass determined by \cite{shore13} ($\le6\times10^{-5}$\Msun), and indeed
greater than the mass ejected in CN eruptions in general.
We are led to conclude that the free-free-emitting
material is unconnected with the 2012 eruption.
This conclusion is consistent with the fact that the free-free emission
is significantly lower than the NIR emission (see Fig.~\ref{herschel-f2}),
which must originate from the 2012 ejecta.

\subsection{NEOWISE data}

\begin{figure*}
 \includegraphics[width=8cm,keepaspectratio]{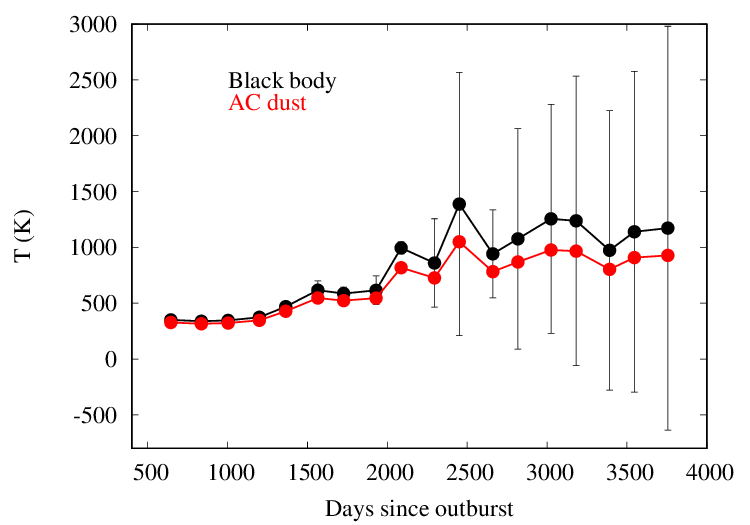}
  \includegraphics[width=8cm,keepaspectratio]{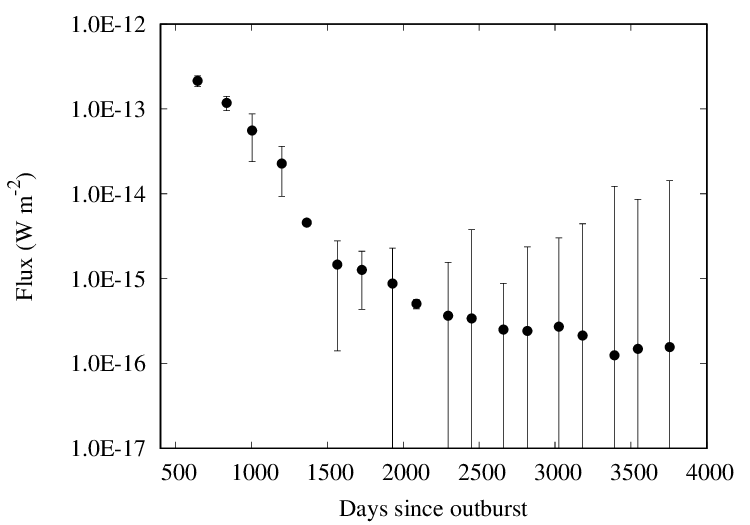}
  \includegraphics[width=8cm,keepaspectratio]{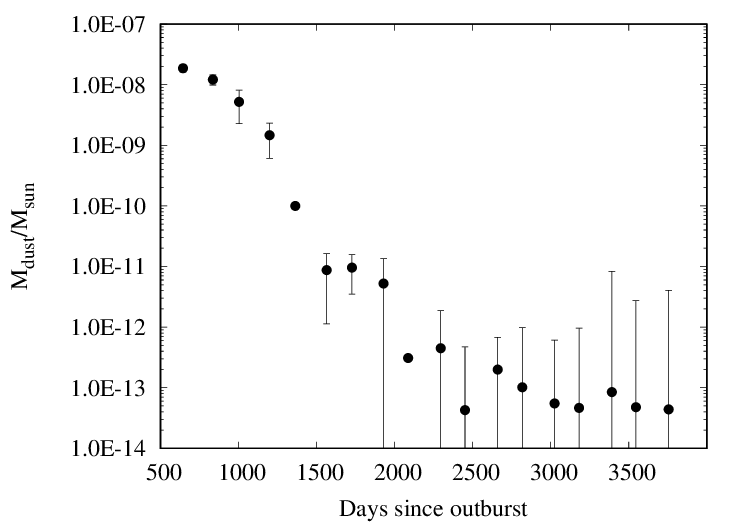}
   \includegraphics[width=8cm,keepaspectratio]{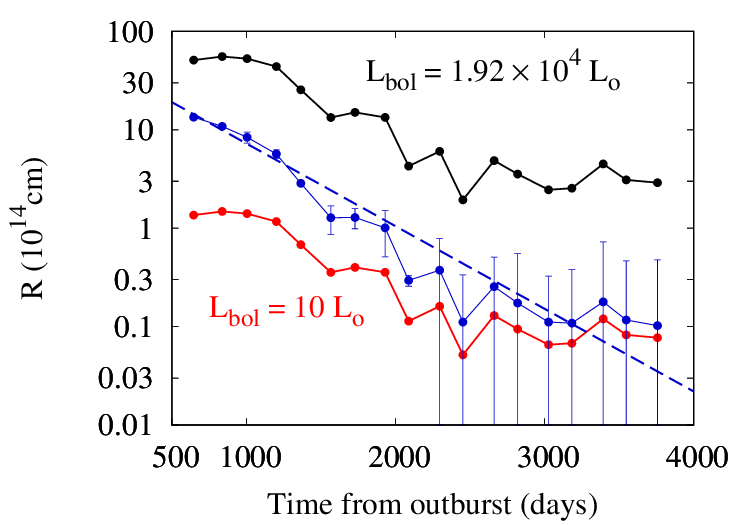}
\caption{Top left: variation of temperature as deduced from
NEOWISE data. Black points are the black body temperatures,
the red points those of AC dust as described in text. Error
bars are for black points. See text for details, and also 
Fig.~\ref{wise_lc} (bottom panel).
 Top right: variation of observed flux from NEOWISE excess.
 Bottom left:  as top right, but for AC dust mass.
 Bottom right: variation of nova--dust distance, assuming that
 the emitting material is AC. Black points are for bolometric
 luminosity $7.35\times10^{30}$~W ($1.92\times10^4$\Lsun), 
 red points for bolometric luminosity 10\Lsun.
 {Blue points are for a bolometric
 luminosity that declines with time, as described in the
 text. The broken curve is a fit of an exponential decline.
 Uncertainties are for blue points only.}
 Lines are included to guide the eye. \label{dust}}
\end{figure*}

The pre-eruption W1 and W2 fluxes have been subtracted from the NEOWISE
data to give the 3.4\mic\ and 4.6\mic\ fluxes from the ejecta.
These excess fluxes have been dereddened by $E(B-V)=0.7$.
The variation of the dereddened flux excesses in NEOWISE bands W1 and
W2 is also shown in Fig.~\ref{wise_lc}, as is the variation of the 
corresponding flux ratio W2/W1. We consider possible origins of the excess flux.

\subsubsection{Emission lines} 
The NEOWISE W1/W2 excess fluxes are unlikely to arise from coronal
(or other emission) lines. Likely 
lines\footnote{{We take these to be those lines that
lie between the 10\% levels of the W1/2 responses.}}
in the NEO\-WISE bands are 
{\fion{Al}{v} 2.905\mic, \fion{Mg}{viii}
3.028\mic, \fion{Ca}{iv} 3.207\mic, \fion{Al}{vi} 3.660\mic,
\fion{Al}{viii} 3.690\mic\ (W1) and
Br$\,\alpha$ 4.052\mic\ and \fion{Mg}{iv} 4.487\mic\ (W2)
\citep[see Figure~1 of][]{evans14}.
The line fluxes (and their variations)
are determined by (i)~the electron temperature and density
and {\em their} variations, and possible abundance 
(and velocity) gradients in the ejecta, and (ii)~the 
critical electron densities, above which the upper levels
of coronal transitions are collisionally rather than radiatively
de-excited. All these factors would have to combine in such a way
that they mimic the monotonic increase in black body temperature
we observe. It is highly unlikely that these would conspire 
to give the variations seen.}

\subsubsection{Free-free emission} 
The flux ratio is such that 
flux(W1) $<$ flux(W2) up to day~1700. Given the presence of optically
thin free-free emission at wavelengths $>70$\mic\ 
(see Section~\ref{herschel}), we consider
whether the NEOWISE data might be consistent with free-free emission. 
We fit equation~(\ref{fff}) to the excess fluxes in the NEO\-WISE data. 
However in general the derived plasma temperatures are $\ltsimeq6000$~K, 
with sporadic outliers $\gtsimeq10^6$~K. We conclude that the 
excess fluxes in the NEOWISE data are not due to free-free emission.

\subsubsection{Dust\label{dustvel}} 
We next consider whether the excess flux might be 
due to dust emission. A simple black body 
\[ f_\nu = \frac{F}{\lambda^3}  \:\: \frac{1}{\exp(B/\lambda) - 1 } \]
has been fitted to the 
flux excesses to give a black body temperature $T$; 
here $F$ is a scaling factor and $B=hc/kT$. The variation
in $T$ is shown in Fig.~\ref{dust}. 
Since we are determining
two parameters ($B$ and $F$) from just two data points, the
uncertainties in $B$ and $F$ were estimated by fitting black
bodies to the maximum and minimum fluxes consistent with
the uncertainties in the flux excess; these uncertainties are
included in Fig.~\ref{dust}. The temperature uncertainties
become very large $\gtsimeq2500$~days from eruption
{because (a)~the values of the W1 and
W2 fluxes approach the faint quiescent values and (b)~of 
the way in which they have been calculated.} 
The bottom panel of Fig.~\ref{wise_lc},
in which the uncertainties in W2/W1 are estimated from the 
uncertainties in the individual fluxes, suggests that (a)~the 
temperature does indeed attain $\sim1300$~K at the latest
times, (b)~the dust temperature increases monotonically and 
(c)~the short-term variations in $T$ after 
$\sim2500$~days are not real.
{Also included in the bottom panel of 
Fig.~\ref{wise_lc} is the flux ratio in the 
Rayleigh-Jeans limit ($\mbox{W2/W1}\simeq0.546$).
The fact that some of the observed ratios at later
times are consistent with this limit indicates that the 
latest temperatures should likely be regarded as
{\em lower} limits.}
The temperature rises from $\sim370$~K to
$\sim1300$~K over the period of the NEOWISE data. 
{Dust emission therefore seems to be the 
most reasonable interpretation of the excess fluxes.}

Having established that the excess flux is due to 
emission by dust, we can place our interpretation of 
the data on a more robust footing by assuming a specific 
grain type. We assume amorphous carbon (AC), 
for which the Planck mean absorption efficiency has the 
particularly simple form
$\langle{Q}_{\rm abs}\rangle = AaT^{\beta}$, where $A = 58.16$ 
for grain radius $a$ in cm 
{(we take $a=0.1$\mic),} and $\beta=0.754$
\citep{evans17}. The resulting AC dust 
temperatures are also included in Fig.~\ref{dust}. 

We estimate the mass of emitting dust using the formulae in
\cite{evans17}. The decline in the dust flux, and that of the 
corresponding dust mass (again assuming AC dust), is shown in
Fig.~\ref{dust}. The mass of dust $M_{\rm d}$ declines from an
initial value $\sim2\times10^{-8}$\Msun\ to $\sim10^{-12}$\Msun\
at later times, 
{although values later than day $\sim2500$
should be regarded as upper limits}.
The initial mass may have been higher, but there are no NEOWISE data before day 644 to investigate this possibility.
{Note that the observed
flux and its variation, together with that of the deduced
dust mass, are independent of the location of the dust in
the \mon\ system.}

{The star--dust} distance $R$ may be
estimated from 
\begin{equation}
 R = \left [ \frac{L_{\rm bol}}{16\pi{a}A\sigma{T}^{(\beta+4)}} 
 \right ]^{1/2} \:\:, \label{RR} 
\end{equation}
where $\sigma$ is the Stefan-Boltzmann constant.
During eruption, CNe essentially maintain constant
bolometric
luminosity, but {in the case of \mon,
the super-soft phase, which traces the TNR and the constant
bolometric luminosity phase, started to decline
on day~200 \citep{page13b} and had essentially 
``shut down'' by day 247 \citep{page13a}. In Fig.~\ref{dust}, 
we first show $R$ for two
extreme cases: (a)~the constant bolometric luminosity
displayed by \mon\ during the super-soft phase, which we
take to be $L_{\rm bol}=7.35\times10^{30}~\mbox{W}$ 
\citep[$1.92\times10^4$\Lsun; based on the day~152 value 
from][scaled to 3~kpc]{shore13} and 
(b)~$L_{\rm bol} =10$\Lsun, appropriate for a CN in
its post-eruption phase. The actual luminosity must lie
between these two extremes.}

{We can disregard the post-eruption 
luminosity ($L_{\rm bol}=10$\Lsun) case, for which
the initial (day~644)} radius is $\sim1.4\times10^{14}$~cm, 
eventually declining to $\sim10^{13}$~cm. This range is well
within the ejecta radius as determined by \cite{shore13}. The
ejecta would have reached this material only $\sim8$~days after
eruption; this would surely have had observable consequences. 

{To explore the declining bolometric
luminosity case, we presume that it declines as described by
\cite{prialnik86}, who found that, after the constant
bolometric luminosity phase, the bolometric
luminosity (for the case of a CO WD) varies approximately as 
$L_{\rm bol} \propto t^{-1.14}$. For \mon, we suppose that 
$L_{\rm bol}$ varies as
\begin{eqnarray*}
 L_{\rm bol} & = & L_{\rm CBL} ~~~~~~~ (0\ltsimeq{t}<{t}_{\rm ss}) \\
  & = & L_{\rm CBL} (t/t_{\rm ss})^{-1.14} ~~~ (t{\ge}t_{\rm ss})
\end{eqnarray*}
where $L_{\rm CBL}=7.35\times10^{30}$~W and $t_{\rm ss}=200$~days 
marks the end of the  super-soft phase.}

{The resultant variation of $R$ is shown as the blue
curve in Fig.~\ref{dust}. The decline in $R$ 
with time is again evident, and is reasonably described by
\begin{equation}
 R = R_{0} \:\: \exp[-\alpha{t}] \:\:, \label{Rt}
\end{equation}
where $R_{0} ~~ (\simeq5.00\times10^{15}$~cm) and
$\alpha ~~ (\simeq1.93\times10^{-3}$~day$^{-1}$; 
note that, while this gives a reasonable description of 
the behaviour of $R(t)$, there is no physical basis for
the exponential dependence).}
{Note also that the light travel
time across the dust shell, $R_0/c\simeq2$~days, 
so ``infrared echo'' effects are negligible
over the timescale of the NEOWISE data ($\gtsimeq500$~days).}

{At $t=200$~days,
$R_0\simeq5\times10^{15}$~cm, which corresponds
to an angular diameter of $0\farcs22$ at 3~kpc;
ejecta moving at 2000\vunit\ would encounter this
material $\sim290$~days after the eruption.}

This suggests that the dust is heated by 
{the stellar remnant as its bolometric
luminosity declined after
the 2012 eruption,} and is located at the distance shown
{by the blue curve} in Fig.~\ref{dust}. 
The dust-bearing material is clearly unconnected with the 2012 
eruption, and must predate this event. 
{This pre-existing material may coincide 
with the ``waist'' of the structure identified in the 
day~882 HST image of \mon\ \citep{sokoloski16}.}
If this pre-existing dust were heated by the 
pre-nova binary, its temperature would have been $\sim70$~K,
and would have had 
$[\lambda{f}_{\lambda}]_{\rm max}\sim8\times10^{-19}$~W~m$^{-2}$
if the nova progenitor had bolometric luminosity 10\Lsun.
It would have been undetectable in the WISE or NEOWISE surveys,
which, as discussed above, revealed only the secondary star.
{On the other hand, 0.1\mic\ AC grains at
$R\sim5\times10^{15}$~cm from a source of radiation
that declines as discussed above would have temperature 
$\sim280$~K on day~644, comparable with that observed in the
earliest NEOWISE data (see Fig.~\ref{dust}).}

{The decline in grain mass (which, 
as already noted, occurs independently of the location of the
dust) presumably arises 
because the dust is destroyed. This could occur by}
{(a)~the 
destruction of ``static'' pre-exiting dust, at constant
distance from the star, or (b)~dust that falls towards the star
and is destroyed as it does so. To explore (a), we rearrange
Equation~(\ref{RR}) to obtain
\[   \left ( \frac{a}{\mic} \right ) = 
\frac{1.76\times10^{11}}{T^{\beta+4}}
\left ( \frac{t}{200~\mbox{d}} \right )^{-1.14} 
\left ( \frac{R}{5.00\times10^{15}~\mbox{cm}}  \right )^{-2} \:\:, \]
where we have used the value of $R$ at 200~days as the fiducial
and the \cite{prialnik86} form for the decline in bolometric
luminosity. We find that the grain radius becomes
less than the dimensions of a C$_2$~molecule 
around day $\sim2000$. However there is obvious dust emission
at least until day~2500, rendering this scenario unlikely.
The implication is that the dust must be falling 
towards the central binary. The velocity with which it does
so is (from equation~(\ref{Rt}))
\[ V_{\rm fall} \simeq 1120 \:\: 
\exp[-0.00193~t]  \mbox{~~\vunit~~,}  \]
where $t$ is in days. This greatly exceeds the free-fall
velocity at $R_0$, which is $\sim3$\vunit.}

\section{Origin of the dust}

The dust interpretation of the flux excess is not straight-forward.
The temperature increase in particular is counter-intuitive.
It is inconsistent with grain formation, in which the starting 
temperature 
would be $\gtsimeq1000$~K, monotonically declining thereafter.
Note that dust formation and subsequent cooling,
as normally occurs in a dust-forming CN, would result in W2/W1 
{\em increasing} with time, contrary to what is observed (see
Fig.~\ref{wise_lc}).

Furthermore, had \mon\ undergone an episode of dust formation
\citep[see, e.g.,][for grain formation in CNe]{evans12}, its 
speed class
would lead us to expect that it would have done so $
\ltsimeq60$~days after the 
2012 eruption \citep[see Figure~2 of][]{williams13}.

\begin{figure*}
 \includegraphics[width=8cm,keepaspectratio]
 {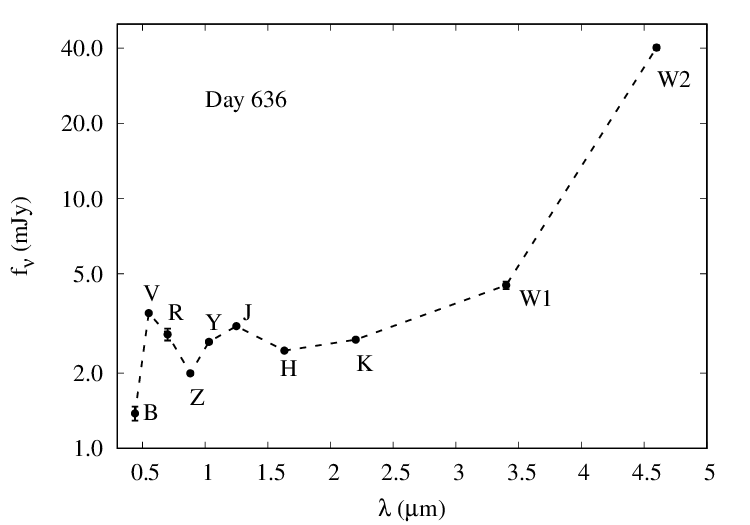}
  \includegraphics[width=8cm,keepaspectratio]
  {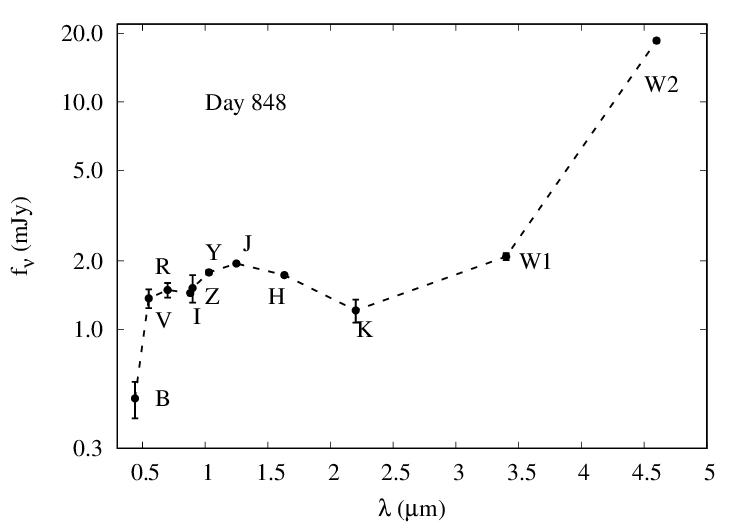}
 \caption{SED of \mon\ around days 636 (left) and 848 (right), 
 from AAVSO, UKIRT WFCAM and NEOWISE data.
 Data as observed (i.e. not dereddened). \label{old-dust}}
\end{figure*}

Moreover, clear evidence that we are not seeing freshly-formed 
dust comes
by combining AAVSO and UKIRT WFCAM data obtained nearly 
contemporaneously with
NEOWISE data. There is sufficient overlap between these datasets 
around days
636 and  848 (see Fig.~\ref{old-dust}; there are a few days' 
offset between the NEOWISE, WFCAM and AAVSO
data). Any freshly-formed dust would be
visible in the WFCAM bands, particularly $K$. 
Moreover, inspection of Fig.~\ref{LC} shows that there was never 
an upturn in the $K$-band light curve 
\citep[see, e.g.,][for an illustration of the 
rise in the $K$ flux when dust forms in CNe]{banerjee16,raj17}.
There is clearly no hot dust.

An IR echo is a possibility, if the echo from more distant (from the nova,
i.e., cooler) dust reaches the observer before that from nearer (to the 
nova, i.e., hotter) dust. A suitable dust geometry could give rise to 
(a)~the cool dust being seen first, followed (gradually) by the hotter dust,
and (b)~the apparent decline in $R$. A dust density gradient such 
that the dust density increases with increasing distance from the nova 
would account for the apparent decline in $M_{\rm d}$. 
However such a scenario 
seems somewhat contrived.

A more plausible interpretation is that there was a dust 
shell around the \mon\ system that pre-dated the 2012 eruption. 
There are two possible origins for such pre-existing circumbinary material:
\begin{enumerate}
\item material left over from the common envelope phase through 
which the CN progenitor evolved. Numerical simulations
\cite[see, e.g.,][]{gonzalez22} suggest that the common envelope ejected
during this phase would have mass $\sim0.1$\Msun;
 \item  \cite{williams08} have argued that 
the transient heavy element absorption (THEA) systems seen in many CNe 
around maximum light indicate the presence of circumbinary gas
that precedes the nova eruption. This material, which likely arises as a 
result of mass ejection from the secondary before  the eruption, 
would have a mass 
$\sim10^{-5}$\Msun, and be located $\sim10-100$~AU 
($\sim1.5\times10^{14}-1.5\times10^{15}$~cm) from the central binary.
\end{enumerate}
In both cases the material would 
expected to be concentrated largely in the binary plane.
However the material expected on the THEA system interpretation
is expected to be much closer to the binary than the 
$R\simeq5.0\times10^{15}$~cm observed, so the common envelope
seems the most likely interpretation.

The mass of the free-free-emitting gas 
detected in the Herschel data is given by equation~(\ref{m-ff}).
This gas, located in the circumbinary envelope,
was flash-ionised by the 2012 eruption. We suggest that both 
the gas identified with the Herschel free-free emission, and 
the dust detected in the NEOWISE data, are associated with 
this pre-exisiting circumbinary material.

The mass of the common envelope is $\sim0.1$\Msun. From equation~(\ref{m-ff}), the implied electron density for 
$T_{\rm gas}=10^4$~K is $n_e\simeq3.24\times10^6$~cm$^{-3}$,
{and $\simeq3.19\times10^7$~cm$^{-3}$ for 
$10^5$~K}. At $10^4$~K and $10^5$~K, the recombination 
time-scale is $\sim20$~days and $\sim10$~days 
respectively, so the ionisation is presumably 
maintained by the hot WD. These values of $n_e$, together with
the common envelope mass, yields an emitting volume 
$\sim2.5\times10^{49}$~cm$^3$ for $10^4$~K,
{and $\sim2.6\times10^{48}$~cm$^3$ for $10^5$~K,}
assuming solar composition. If the common envelope is in
the form of a torus with major radius $R\sim5.0\times10^{15}$~cm
and minor radius $\eta{R}$ $(\eta<1)$,
the volume is $\sim2.5\times10^{48}\eta^2$~cm$^{3}$. Given the 
approximations made, these estimates for the envelope volume
are sufficiently close to 
suggest that this is a plausible interpretation.
Further, if the (initial) $\sim2\times10^{-8}$\Msun\ of dust is uniformly
distributed throughout the torus, there are 
$\sim2\times10^{-9}$ 0.1\mic\ AC grains cm$^{-3}$,
which leads to an optical depth in the visual 
$\sim6\times10^{-3}\eta^{-1}$ through the thickest part of the
torus. The dust in the torus has essentially 
no extinction effect, 
{even if its axis is perpendicular to
the line of sight.}

\section{A possible interpretation}

We propose a qualitative interpretation in which the ejecta from the 
2012 eruption encounter the pre-exisiting circumbinary material ejected
during the nova progenitor's common envelope phase. We suggest 
that, in \mon, the density of the latter exceeds that of the 
former where the two materials interact.
We recall our estimate that the 2012 ejecta would reach the pre-exisiting
dust shell in $\sim290$~days which, given (a)~the approximate nature of
our estimates and (b)~the fact that there are no NEOWISE data prior to
day~644, is consistent with the time that the IR excess is first seen in the 
NEOWISE data. This interaction causes the ejecta to decelerate, 
producing conditions that are conducive to the formation of 
Rayleigh-Taylor instabilities. As a result the denser (dust-containing)
material from the circumbinary material forms ``blobs'' and falls
``downwards'' towards the central binary, while the less dense ejecta 
accelerate into, and penetrate, the circumbinary material.
{The amount of dust seen in \mon\ 
($<10^{-8}$\Msun) is orders of magnitude less than
that expected to have formed in a common envelope 
\citep[$\sim2\times10^{-8}$\Msun; see, e.g.,][]{iaconi20}.
As discussed in Section~\ref{dustvel}, the dust falls at 
velocity $\sim1120\,\exp[-0.00193\,t]$\vunit\ 
($t$ in days), greatly in excess of the free-fall velocity
at the distance of the torus. The free-fall timescale 
would be $\sim10^3$~years rather than the $\sim10^3$~days
observed. We speculate that the dust may be carried inwards by a
reverse shock, similar to that seen in 
the supernova remnant Cassiopeia~A \citep*{vink22}.}

As the dust falls, its temperature increases even as the 
bolometric luminosity of the central source declines, 
as observed (see Fig.~\ref{dust}). 
The declining dust mass must be due to destruction of the dust.
This can not be due to evaporation, as the observed temperatures 
start at $\sim350$~K and remain well
below the evaporation temperature for AC \citep[$\sim1800$~K;][]{ebel00}.

``Physical'' sputtering, in which atoms are removed from the grain surface
by the impact of a gas atom or ion, can not be important, because the 
threshold for removing an atom from the grain surface is $\gtsimeq10$~eV,
greater than the typical thermal energy of a gas atom/ion at
$10^4$~K, $\sim1$~eV. 

Chemisputtering,
in which the impacting atom/ion bonds with a surface atom and removes it,
essentially has no threshold. We propose that, as the dust is exposed to
the ejecta, it is subject to erosion, by chemi\-sputtering 
\citep[see][for an application to CNe]{mitchell84}. 
The rate of erosion is given by 
\[  \dot{a} = - \frac{n_{\rm H}Ym_{\rm C}}{4\rho} \left ( \frac{8kT_{\rm gas}}{\pi{m_{\rm H}}} \right )^{1/2} \:\:, \]
where $n_{\rm H}$ is the hydrogen number density in the gas, $Y$ is the 
chemi\-sputtering yield, and $m_{\rm C}$, $m_{\rm H}$ are the masses 
of the C and H atoms respectively.

The chemisputtering yield for
atomic hydrogen on amorphous hydrogenated carbon, covering the surface
temperature range of interest here, has been measured
in the laboratory by \cite{salonen01}.
The yield at $T\simeq300$~K is $\sim5\times10^{-3}$, rising to a peak of 
$\sim2\times10^{-2}$ at 900~K, then falling back to 
$\sim5\times10^{-3}$ at $\sim1100$~K.
Assuming that the gas number density at 
$1.5\times10^{15}$~cm was $3.2\times10^{7}$~cm$^{-3}$ \citep{shore13},
and that the density declines as $r^{-2}$, 
the hydrogen density at $6\times10^{15}$~cm (see above) was 
$2\times10^{6}$~cm$^{-3}$.

For gas temperature $10^4$~K, a 350~K AC grain erodes at 
$\sim3.1\times10^{-5}$\mic\,day$^{-1}$, while a 900~K grain 
erodes at $\sim1.3\times10^{-4}$\mic\,day$^{-1}$.
From Figs~\ref{wise_lc} and \ref{dust}, we see that an 
AC grain in the environment of \mon\ had temperature 
$800-1000$~K {after day 2000}. During this time chemical erosion would have been
at its most potent. The lifetime of a 0.1\mic\ grain at 
the start of the NEOWISE observations would have been $\sim3000$~days,
falling to $\sim800$~days as the grains became hotter. Chemisputtering 
is able to all but destroy the grains by the end of the NEOWISE
observations, and is more than adequate to account for the 
decline of dust mass. Moreover, as the grains decrease in 
size, the effects of \citep[by this time diminishing; see][]{page13a} 
X-ray and ultra-violet radiation on the dust 
\citep*[e.g., by grain charging and subsequent destruction
by electrostatic stress;][]{fruchter01} might come into play, 
thus accelerating the destruction process.

Whether or not this phenomenon occurs in CN eruption in general
depends on the relative densities of the ejecta and the common envelope
material they encounter. Also, if the grains in this material were
silicate rather than AC, they would not be subject
to the destructive effects of chemisputtering. We propose that the 
NEOWISE archive be interrogated for similar behaviour in the 
aftermath of future CNe eruptions.

\section{Conclusion}
NEOWISE data for \mon\ suggest that dust was present in the nova system
prior to the 2012 eruption. As the ejecta from the eruption encountered
the pre-existing dust-bearing material, Rayleigh-Taylor instabilities
gave rise to dust clumps which rained down onto the central binary.
The relative densities of the pre-existing material and ejecta determine
whether this phenomenon occurs in nova eruptions in general.

\subsection*{Acknowledgements}

{We thank the referee, Professor Bob Gehrz, 
for his very helpful comments on an earlier version of this paper.}

This publication makes use of data products from the Wide-field Infrared 
Survey Explorer, which is a joint project of the University of California, 
Los Angeles, and the Jet Propulsion Laboratory/California Institute of 
Technology, funded by the National Aeronautics and Space Administration.
It also makes use of data products from NEOWISE, which is a project of 
the Jet Propulsion Laboratory/California Institute of Technology, funded 
by the Planetary Science Division of the National Aeronautics and Space 
Administration.

This work has made use of data from the European Space Agency (ESA) mission
{\it Gaia} ({https://www.cosmos.esa.int/gaia}), processed by the {\it Gaia}
Data Processing and Analysis Consortium (DPAC,
{https://www.cosmos.esa.int/web/gaia/dpac/consortium}). Funding for the DPAC
has been provided by national institutions, in particular the institutions
participating in the {\it Gaia} Multilateral Agreement.

UKIRT is currently owned by the University of Hawai'i (UH) and operated
by the UH Institute for Astronomy. Over the period 2014--2017, 
UKIRT was supported by NASA and operated under an agreement among the UH,  
the University of Arizona, and Lockheed Martin Advanced Technology Center;
operations were enabled through the co-operation of the East Asian Observatory.
During the period of the earlier observations, UKIRT was operated by the Joint
Astronomy Centre on behalf of the Science and Technology Facilities Council
of the UK. We thank the UKIRT staff for carrying out the observations and the
Cambridge Astronomy Survey Unit for processing the data.

PACS has been developed by a consortium of institutes led by MPE (Germany) 
and including UVIE (Austria); KU Leuven, CSL, IMEC (Belgium); CEA, LAM 
(France); MPIA (Germany); INAF-IFSI/OAA/OAP/OAT, LENS, SISSA (Italy); 
IAC (Spain). This development has been supported by the funding agencies BMVIT
(Austria), ESA-PRODEX (Belgium), CEA/CNES (France), DLR (Germany), 
ASI/INAF (Italy), and CICYT/MCYT (Spain).

This paper also makes use of data products from the Two Micron All Sky 
Survey, which is a joint project of the University of Massachusetts and 
the Infrared Processing and Analysis Center/California Institute of 
Technology, funded by the National Aeronautics and Space Administration
and the National Science Foundation.

We acknowledge with thanks the variable star observations from the AAVSO 
International Database contributed by observers worldwide and used in 
this research.

\section*{\bf Data availability}
The data used in this paper are available as follows:
 
\noindent WISE and NEOWISE: https://irsa.ipac.caltech.edu/ \\
cgi-bin/Gator/nph-scan?submit=Select\&projshort=WISE

\noindent  Herschel: https://irsa.ipac.caltech.edu/applications/Herschel/

\noindent  The UKIRT data are available on-line.

\bsp	
\label{lastpage}
\end{document}